\documentclass[12pt,preprint]{aastex}
\usepackage{psfig,color}

\shorttitle{Chandra observations of IM Nor }
\shortauthors{Orio et al.}

\begin{document}


\title{Chandra observations of the recurrent nova IM Nor
}


\author{Marina Orio}
\affil{INAF, Osservatorio Astronomico di Torino, Strada Osservatorio, 20,
I-10025 Pino Torinese (TO), Italy\\
and Department of Astronomy, 475 N. Charter Str. University of Wisconsin, 
Madison WI 53706}
\email{orio@astro.wisc.edu}

\author{Emre Tepedelenlioglu}
\affil{Department of Physics, University of Wisconsin,
    1150 University Avenue, Madison WI 53706}
\email{emre@cow.physics.wisc.edu}

\author{Sumner Starrfield}
\affil{Department of Physics and Astronomy, Arizona State University,
Tempe, AZ 85287-1504}
\email{starrfield@asu.edu}

\author{Charles E. Woodward} 
\affil{Department of Astronomy, University of Minnesota, 
116 Church Street, SE, Minneapolis, MN 55455}
\email{chelsea@astro.umn.edu}
\and

\author{Massimo Della Valle}
\affil{INAF, Osservatorio Astrofisico di Arcetri, largo E. Fermi 5, 50125 Firenze,
 Italy}
\email{massimo@arcetri.astro.it}



\begin{abstract}
The recurrent nova IM Nor was observed twice 
in X-rays with Chandra {\it ACIS-S},
1 and 6 months after the optical outburst.
It was not detected in the
first observation, with an upper limit on the X-ray luminosity 
in the 0.2-10 keV  range   L$_{\rm x} < 4.8  \times 10^{30}
\times$ (d/1 kpc)$^2$  erg s$^{-1}$ (where d is the
distance to the nova).
Five months later,  a hard X-ray source 
with L$_{\rm x} = (1.4-2.5) \times 10^{32}
\times$ (d/1 kpc)$^2$ erg s$^{-1}$ was detected. 
The X-ray spectrum appears to be thermal, but we cannot rule out
 additional components due to unresolved emission lines.
 A blackbody component is likely to contribute to the observed
 spectrum, but it has bolometric luminosity
L$_{\rm bol} =  2.5 \times  10^{33} \times$ (d/1 kpc)$^2$
 erg s$^{-1}$, therefore it is not sufficiently luminous to be
due to a central white dwarf which is still burning
hydrogen on the surface.  An  optical spectrum, taken 
5 months post-outburst, indicates no intrinsic reddening of
the ejecta.  Therefore, we conclude that the shell had
already become optically thin to supersoft X-rays, but nuclear burning had 
already turned off,  or was in the process of turning off
at this time. We discuss why this implies that recurrent novae, even the
 rare ones with long optical decays like IM Nor, indicating a large envelope mass,
are not statistically significant as type Ia SN candidates. 
 
\end{abstract}

\keywords{binaries: close--novae, cataclysmic variables -- 
stars: individual(IM Normae) --- X-rays: stars}

\section{Introduction}

IM Nor is a {\it Recurrent Nova} (RN),
that was observed in outburst in 1920 and again on 2002 January 3
(Elliot \& Liller 1972, Liller 2002a).
 Assuming that the maximum occurred on  2002 January 14,  
at  V=7.84  (Liller 2002b; note that a successive measurement
 V=7.66 on January 16 was done through fog
and had a large error bar, Liller 2002c),
the times to decay by 2 and 3 magnitudes, respectively, were
in the ranges
t$_2$=21-26 days and t$_3$=45-74  days  (Pearce 2002, Shida 2002). 
An approximate value t$_3 \approx$ 50 days
is derived  from the VSNET light curve by Kato et al. (2002).
These times  are  longer than for most RN,
with the exceptions of T Pyx (Webbink et al. 1986 and
 references therein) and CI Aql (Kiss et al. 2001).
 The FWHM of the
emission lines was 1150 km s$^{-1}$, indicating
moderately high ejection velocities (D\"urbeck et al.
2002). IM Nor is the only known
RN classified as a  ``Fe II nova'' (see Retter et al.
2002, D\"urbeck et al. 2002 and references therein).   
Another peculiarity is that IM Nor  may have ejected significantly
more mass than all other RN, except CI Aql (see Retter et al.
2002, and discussion below).

   Classical novae are thought to be recurrent over time scales
$\geq$1000 years, while RN undergo inter-outburst periods of
only $<$100 years. The reason seems to be due to
 a more massive white dwarf (hereafter, WD) than in the average 
classical nova system, and in some cases higher mass accretion rate 
(see early work by Fujimoto 1982 and
Prialnik et al. 1982, for the dependence
of inter-outburst time on physical parameters of the system). 
RN are interesting as possible progenitors
of type Ia Supernovae. A list of
 publications on the subjects includes Starrfield et al.
 (1985, 1988),   Livio \& Truran (1994), Della Valle
 \& Livio (1996), Hachisu \& Kato (2000). There are basically two
types of RN, long period systems hosting a red giant, and
 typical cataclysmic variables with only a slightly evolved
secondary. For the latter group, orbital periods seem to be either
 very short compared
to other novae (2-3 hours), or quite long (10-24 hours).
 IM Nor has a short
orbital period,  2.46 hours (Woudt \& Warner 2003)

Novae at quiescence are X-ray
 sources because of accretion, but they are faint (Orio
 et al. 2002). After an outburst,
they turn into  more luminous X-ray sources,  initially
because of shocks in the ejected shell
(e.g., in  colliding winds,
or phenomena between ejecta and circumstellar matter).
Thermal bremsstrahlung emission is detected
at temperatures in the range
0.1-10 keV, and at late phases emission lines have  also been
detected in the 0.2-1.0 keV
energy range. The X-ray luminosity in the 0.2-10 keV range is
 L$_{\rm x}=$10$^{33-35}$ erg s$^{-1}$ (see Orio 2004 for a review).
If the  central WD keeps on burning un-ejected
hydrogen, which has  
 interesting implications related to the final fate of nova systems,
then novae appear  as  luminous supersoft X-ray sources with
bolometric luminosities  in the range  
L$_{\rm x}= $10$^{36-38}$ erg s$^{-1}$. The WD atmosphere is extremely hot, 
(T$_{\rm eff} \geq 2 \times 10^{5}$ K), 
and directly observable when the ejecta become
optically thin to supersoft X-rays. 

In this case, the atmospheric
absorption edges and  many  narrow and crowded absorption
lines are detected and, for the most luminous objects, 
can be resolved with grating observations in X-rays (e.g., Ness et al. 2003).
In the Galaxy only  $\approx$20\% of all classical and
recurrent novae were observed as supersoft X-ray sources
 for more than a few months (see Orio 2004).
The length of the supersoft X-ray phase is an indication of the amount
of hydrogen fuel left over after each outburst,
hence the likelihood that the white dwarf mass grows towards
 the Chandrasekhar limit. 
 The nature of type Ia SN progenitors is still an open
question, but  single degenerate close binary systems,
and among them especially RN,
are thought to contribute significantly to the type Ia SN rate (e.g. 
Starrfield et al. 2004).
 If not all the accreted material is ejected and the WD mass grows
after repeated outbursts, RN can be considered the most likely
progenitors of type Ia SN among novae and CV (it
is crucial however to estimate the abundances,
because one needs to rule out the possibility that eroded WD 
material, instead of freshly accreted material, is being burned). 

 Della Valle \& Livio                         
(1996) argued that the specific contribution of RN to 
the type Ia SN rate is not  high. 
The discovery of IM Nor and 
of another  RN, CI Aql, in outburst in 2001,  
may change these conclusions. A long decay time
is thought to indicate a large envelope mass.  In both these two unusual RN, 
the empirical relationship between t$_2$ and and ejected mass $\Delta$m$_{\rm ej}$
of Della Valle et al. (2002) yields 
$\Delta$m$_{\rm ej}$ of order of 10$^{-4}$ M$_{\odot}$.  
Theoretically, Hachisu et al. (2003) estimate
instead an envelope mass $\Delta$m$_{\rm env}$=8 $\times 10^{-6}$ M$_{\odot}$
for CI Aql. In the models,  for reasons that
are not clear yet, an order of magnitude
smaller $\Delta$m$_{\rm ej}$ is commonly predicted than it is inferred
from the observations (e.g. Gehrz et al. 1998).
However, even this lower 
theoretical value is much higher than those estimated for most other RN.
If the retained
mass after the outburst was a significant fraction of the
ejected amount, then these two objects may be representative of a class of RN 
which accumulate 
enough material on the WD despite ejection during outbursts, 
to grow towards the Chandrasekhar mass in a time that
 is sufficiently short to have an impact on the type Ia SN rate.
 This class probably includes 
T Pyx, whose outbursts do not seem to be recurrent any more on a fixed time
scale, but are also very long-lasting.  For comparison
U Sco, a  non-red giant RN system in which the decline of the optical
light curve was instead very rapid, ejected only about 10$^{-7}$ M$_\odot$\
(Williams et al. 1981).   
 U Sco appeared as a supersoft X-ray
 source already 20 days after the outburst (Kahabka et al.
1999). This rapid evolution indicates a massive white dwarf, which ejects
little accreted mass, but still retains part of it after
the outburst. U Sco was not observed again in X-rays, so there is no estimate
of the total amount of retained mass. Optical studies suggest, however, that
 only a small amount of mass was ejected, just what is required
 to trigger the nova explosion on a massive WD (Starrfield et al. 1985).
If part of the envelope is retained after RN outbursts, then the
RN with unusually large envelope masses are of great interest for type Ia
SN theories because their WD
may grow rapidly to reach the Chandrasekhar mass. 
 
 Despite the unusually large optical luminosity,
 which is thought to indicate ongoing thermonuclear burning, T Pyx has not been
detected as a supersoft X-ray source (Greiner et al. 2003),
but it was observed only about 30 years after the last outburst. 
 Since the optical light curve of CI Aql evolved slowly (t$_2$=30$\pm$1 days
and t$_3$=36$\pm$1 days, Kiss et al. 2001), X-ray observations 
  were done only more than a year after the outburst, with the
 working hypothesis that the evolution in X-rays would be as slow as the
optical, and that H-burning would also last longer. 
However, X-ray emission was only 
detected from the ejecta and not from the central source (Greiner
 and Di Stefano 2002).

\section{The first Chandra observation}

We asked to perform {\sl Chandra ACIS-S}  X-ray observations of IM Nor
at much earlier epochs than those done for CI Aql, 
 approximately 1 and 6 months post-outburst. The first observation
 was done during the DDT (Director's Discretionary Time) 
with the comparison with U Sco in mind. 
The second observation was done in the framework of an accepted 
Target of Opportunity (ToO) program,
based on the optical spectrum having become nebular (see Section 3).

The date of the first observation was 2002 February 12, almost 4 weeks
after optical maximum, with an  exposure of 5630 s. The data were analysed
with the {\sl CIAO} software for {\sl Chandra} data analysis.
One month post-outburst U Sco had already became an X-ray supersoft
source, and all observed novae were X-ray sources due to
 shell emission (see Orio 2004).  IM Nor however was not detected
in this early observation, with an upper limit
on the {\sl ACIS-S} count rate of 0.0014 cts s$^{-1}$ 
in the 0.2-10 keV range. The value of the equivalent column density
of neutral hydrogen, N(H) in the direction of the nova
obtained from the {\sl NH} program included in NASA's
{\sl FTOOLS} (based on the maps of Dickey \& Lockman, 1990)
yields an average N(H)=8.15 $\times 10 ^{21}$ cm$^{-2}$
in a cone of 1$\rm ^o$ radius in the direction of the
nova, although there is a rather large uncertainty. 
Assuming a blackbody at T=30 eV (e.g., Orio et al.
1999), and conservatively a rather large value N(H)=10$^{22}$ cm$^{-2}$
(which would imply some intrinsic absorption of the ejecta)
this limit in the count rate translates into an upper limit to
the X-ray flux at 0.2-10 keV, 
F$_{\rm x}<$10$^{-14}$ erg cm$^{-2}$ s$^{-1}$. The flux upper limit
is instead 
  F$_{\rm x}<4 \times$ 10$^{-14}$ erg cm$^{-2}$ s$^{-1}$, assuming
instead  a thermal plasma at kT=3 keV and N(H)=10$^{21}$ cm$^{-2}$
(appropriate for for nebular X-ray emission after a month,
 e.g., Mukai \& Ishida 2001). These fluxes correspond to X-ray
luminosity L$_{\rm x}$ = 1.2 and 4.8 $\times 10^{32}
\times$ (d/1 kpc)$^2$ erg s$^{-1}$, respectively.   

\section{Caveats from the optical spectrum}

Optical spectra of IM Nor were obtained on 2002 April 29.40~UT on
the NOAO CTIO Blanco 4-m
telescope using the Ritchey-Chretien spectrograph ($f/7.8$) with
the Blue Air Schmidt camera and the Loral thinned 3K $\times$ 1K format CCD
(Woodward et al. 2002).  A 527 line mm$^{-1}$ grating (KPGL3) resulted in a
dispersion of 1.91 \AA \ pixel$^{-1}$. All spectra were obtained with a
1\farcs3 wide slit and a UV (WG360; 3600 \AA) blocking filter to
reduce second-order contamination in the red. Complete details of
observational and reduction techniques can be found in
Skillman et al. (2003). One of three spectra is shown in Fig. 1.

\subsection{The reddening}

 Due to the contamination of the H $\alpha$ line by the [N II] line,
 and to the contamination of H $\gamma$ by the [O III] line,
we can only estimate the value of E(B--V) with very large
uncertainty. Measuring the equivalent width of the Na I line at 5890 \AA \
yields an approximate value E(B--V)$\approx$1.1 using the empirical
 relationship of Della Valle \& Duerbeck (1993), but we caution
that another  empirical relationship given by Barbon et al.
(1990) yields only E(B--V)$\approx$0.50. Following Ryter et al. (1975)
we obtain N(H)$\approx 7.4 \times 10^{21}$ cm$^{-2}$
in the first case, and only N(H)$\approx 3.4 \times 10^{21}$ cm$^{-2}$
in the second. In any case, these values do not exceed
the one indicated from Dickey \& Lockman
(1990), and therefore we do not find evidence of
intrinsic nebular reddening. This conclusion is definitely
strengthened  by the  ratio of H$\delta$/H$\beta$, which
is known to be 0.26 in stationary conditions (when this ratio depends only on
the recombination ratio).  Intrinsic nebular reddening
 would cause a lower value to be measured. This ratio in our case
 is instead 0.51, which implies that optical depth effect and collisions
changed the recombination ratios, and that we were able to
observe the spectrum where such effects occur. We came to
the conclusion  that  intrinsic
nebular reddening is absent and should not
 be added to the interstellar value.  In other words, the
nebular material was already optically thin at the date of this observation.
      
\subsection{Can we determine the distance?}

The distance to IM Nor is not known, and the distances
to RN in general are difficult to evaluate. For a classical nova, 
 having an estimate of the reddening and of t$_2$ we
should be able to derive the distance from the Maximum Magnitude
versus rate of Decline Relationship (MMRD).  
 However, even if some RN, such as T CrB, RS Oph and V745 Sco,
seem to follow the MMRD, Warner (1995) notes that for T Pyx the MMRD yields 
d=2.1 kpc while the nebular parallax indicates d=1 kpc. Moreover,
observations of RN in M31 (Capaccioli et al. 1989) and LMC (Capaccioli
et al. 1990) show that they are often  1-2 mag fainter than predicted
by the MMRD.  Woudt \& Warner (2003) also
note that the light curves of both CI Aql and IM Nor are different from that of
classical novae and that a correction to the MMRD relationship must
be necessary. Adopting the MMRD of Della Valle \& Livio (1995,  but subtracting
 1 magnitude from the value of the absolute magnitude
at maximum, and using E(B-V)=1.1, we obtain a distance
range 3.8--4. kpc for t$_2$=21-26 d, however with E(B-V)=0.50  
obtained with the Barbon et al. relationship, 
 we obtain  1.5-1.6 kpc.  Since IM Nor was a RN
 with a relatively massive envelope and did not eject as
little mass as U Sco, the correction to the MMRD may
 be even much less than 1 magnitude, and an absolute
 upper limit can be considered d=6.4 kpc, obtained 
without correction to the MMRD.  1.5-6.4 kpc is a large  range, so  
we will continue to parameterize the luminosity as a function
of the distance.

\section{The second Chandra observation}
 
IM Nor was detected in X-rays when it was
observed again for 4930 s  with {\sl ACIS-S} 6 months after the outburst,
on 2002 May 31. We knew that the nebula
was optically thin to supersoft X-rays by this time, but the X-ray source 
 was neither     ``super-soft'' nor very luminous. Further
{\sl LETG} grating observations guaranteed for the TOO program in
 case of a bright nova were not
 scheduled, because only very low S/N would have been achieved.   
The hard   X-ray emission most likely originated in  the ejected nebula
(see Orio 2004 and references therein).  The measured count rate was
0.267$\pm$0.008 cts s$^{-1}$, a factor of 30 higher
than for CI Aql 16 months after the outburst.
 The observations span about
59\% of the orbital period, but  we cannot detect,
nor rule out variability on time scales shorter than the observation time.
 The background
corrected count rate varies only within a 2 $\sigma$ error. 

 In Fig. 2 we show the {\sl Chandra ACIS-S} X-ray spectrum
 and one  possible model fit (see Table 1 and discussion below).
At this stage IM Nor was a harder X-ray source than CI Aql,
which is  expected for an earlier cooling stage of the shell.
We fitted the spectrum with several models available
in the {\sl XSPEC} software package (Arnaud et al. 1996).
In Table 1 we give the main parameters of the spectral fits
and the value obtained for the reduced $\chi^2$.
We let N(H) vary as a free parameter. If
the spectral fit yields N(H)$\leq$ 8.2 $\times$ 10$^{21}$ cm$^{-2}$,
we consider the result consistent with only
interstellar absorption.  The power law model does not give a good fit with
reasonable parameters (we obtain a slope $\alpha$=3.6 and
$\chi^2$/dof=1.5, where dof=degrees of freedom). We also used five 
thermal plasma models included in {\sl XSPEC}. Using only one
component, we obtain a relatively reasonable fit with the
 {\sl bremsstrahlung} model
of {\sl XSPEC}, which  yields  $\chi^2$/dof=1.2.
This fit yields N(H)=2.6 $\times 10^{21}$ cm$^{-2}$, which 
does not indicate intrinsic absorption of the ejecta.
X-ray fluxes  F$_{\rm x}$ and luminosity 
 L$_{\rm x}$ are measured in the 0.2-10 keV range.
The observed flux is F$_{\rm x}$=1.34 $\times$ 10$^{-12}$ erg 
cm$^{-2}$ s$^{-1}$, corresponding to an unabsorbed flux 
F$_{\rm x}$=2.10 $\times$ 10$^{-12}$ erg cm$^{-2}$ s$^{-1}$,
and to L$_{\rm x} = 2.5 \times 10^{32} \times$ (d/1 kpc)$^2$ erg s$^{-1}$.

 One of the models' parameters used in {\sl XSPEC}
is a constant ``N'', which is a function of the distance d to the source.
For the {\sl bremsstrahlung} model 
 N=3.02 $\times$ 10$^{-15}/(4 \pi$ d$^2$ EM),
and for the  {\sl MEKAL}, {\sl VMEKAL} and the {\sl Raymond-Smith}
models,  N=10$^{-14}/(4 \pi$ d$^2$ EM), where 
EM=$\int$ n$_{\rm e}$ n$_{\rm i}$ dV  is the emission measure,
and n$_{\rm e}$ and n$_{\rm i}$ are the electron and ion
density, respectively, integrated over the volume of the nova shell.
In first approximation, we assumed EM=n$_{\rm e}^2$ V, where V is the volume
of the shell. Since the velocity of the
ejecta was around 1000 km s$^{-1}$ (D\"urbeck et al. 2002),
at the date of the observation V$ \simeq 8.75 \times$ 10$^{45}$ cm$^3$. 
 Electron densities
in the ejecta six months after the outburst are known to be  in the range 
10$^5$- few 10$^6$ cm$^{-3}$ (e.g. Bohigas et al. 1989). 
The {\sl  bremsstrahlung}  model
yields EM=6.59 $\times 10^{55}$, implying   a
reasonable value for the electron density, 
n$_{\rm e}\simeq 6.3 \times 10^4$ cm$^{-3} \times$ (d/1 kpc)$^2$,
 if  all of the shell 
is emitting X-rays. We note  
however, that the filling factor was only 
about 1/10th for V382 Vel (Della Valle et al. 2002) and probably 
 much lower for T Pyx (Shara et al. 1997).
 
The {\sl Raymond-Smith}
 model without additional components does not yield a good
fit ($\chi^2$/dof=2.5). With the {\sl MEKAL} model we 
could fit the spectrum only by allowing
the metalicity parameter, Z, to vary. We find
 a reasonable fit with  a very low value,
 Z$\approx$0. To better test the effect of the chemical composition, 
 we tried the {\sl VMEKAL} and the {\sl VRAYMOND}
models, which include
abundances of several elements. We allowed the abundances
of He,C,N,O,Ne,Mg,Ca,Fe and Ni to vary freely. We obtained the best fits
in both models with enhanced C and O, and depleted Fe.
The most outstanding abundances are  
C/C$_\odot$=160, N/N$_\odot$=188, Fe/Fe$_\odot \simeq$0
for {\sl VMEKAL}, and close values were
obtained using {\sl VRAYMOND}. With {\sl VRAYMOND},
the fit is slightly improved with He/He$_\odot$$\leq$0.002, but 
low He abundance  is not required fitting with the {\sl VMEKAL} model. 
These results are surprising (especially for iron, because
IM Nor was a Fe II nova)
 but not conclusive. First of all, 
it is possible to fit the spectrum even better with solar
abundances, by adding an additional component (see below).
 Moreover, there may be non-ionization
equilibrium conditions
(see discussion by Mukai \& Ishida, 2001, for Nova V382 Vel,
for which iron was not detected in the X-ray spectrum).
 
As a next step, we tried fitting composite models.
We added a black-body component to the thermal plasma,
and obtained the best fit of all ($\chi^2$/dof=0.95, see Fig. 1).  
The blackbody fit at T$_{\rm bb}$ = 70 eV is obtained with a constant
proportional to the bolometric luminosity and inversely
proportional to the distance, and in our fit
L$= 2.5 \times  10^{33} \times$ (d/1 kpc)$^2$ erg s$^{-1}$
(note that T$_{\rm bb} \leq 60$ eV
is ruled out by the spectral fits).  This value of the bolometric luminosity,
for any value of the distance,
is orders of magnitudes lower than observed for supersoft X-ray
 sources (see review by Orio, 2004). Thus, a blackbody component
is likely to exist, but we are not observing
the  atmospheric emission of a significant
portion of the surface of a hot, hydrogen burning white dwarf.
The residuals in the fits (see Fig. 1) are suggestive of emission
lines, like those observed previously for Nova V382 Vel (Burwitz et al. 2002).
Additonal narrow Gaussian components
at $\approx$1 and $\approx$1.2 keV, where residuals
 appear in the model fit, do not  fit the data. Adding a
 Gaussian component at 1.94 keV (possibly due to a Ni XVI line) to
the bremsstrahlung model, we improved the
fit to some extent,  although we did not improve it further adding more
Gaussian components. The low spectral resolution prevents
 a very thorough investigation of how emission lines may be altering
the observed spectrum.  All the spectral fits in Table 1 yield 
unabsorbed flux
F$_{\rm x}$=1.2-2.4 $\times$ 10$^{-12}$ erg cm$^{-2}$ s$^{-1}$,
and L$_{\rm x} = 1.4-2.9  \times 10^{32} \times$ (d/1 kpc)$^2$ erg s$^{-1}$. 
 From the values of the emission measure we derive 
  electron density  n$_{\rm e} \approx 10^5-10^6$ cm$^{-3}$ for
d$\approx$ 4 kpc, if all the shell volume is filled.

 We also tried fitting a two component thermal plasma, the last in Table 1, 
which could originate in different
clumps or layers of the ejected nebula and therefore differently absorbed.
Using the {\sl Raymond-Smith} model for both, we find 
a heavily absorbed component at low temperature, 
filling a fraction of at least a 10th of the ejected shell.
The  absorbed flux in this model   
would be 1.8 $\times 10^{-12}$ erg cm$^{-2}$ s$^{-1}$, corresponding to
the high
unabsorbed flux F$_{\rm x}$=6.1 $\times 10^{-11}$  erg cm$^{-2}$ s$^{-1}$.
However, the conclusions we derived from the optical
 spectrum rule out this component
because of the high value N(H)=3.2 $ \times 10^{22}$ cm$^{-2}$.

\section{Conclusions}

The upper limit on the X-ray luminosity of IM Nor a month 
after the outburst is unprecedented, because all classical
and recurrent novae that were observed at early post-outburst epochs 
showed X-ray emission at L$_{\rm x} >  10^{32}$ erg s$^{-1}$
 much earlier than after one month (see review by Orio, 2004).
Half a year after the outburst the nova was detected as an X-ray source. Hard X-ray
emission from the ejecta, with L$_{\rm x} \simeq 
10^{33}$ erg s$^{-1}$,  was the dominant component in
the X-ray spectrum.
 
A blackbody component is also likely present in
the X-ray spectrum of IM Nor, at high temperature,
T$_{\rm bb} \simeq$ 70 eV, but at such low luminosity,
L$_{\rm x} \leq 2.5 \times 10^{33} \times$ (d/1 kpc)$^2$
erg s$^{-1}$, that does not 
fit the observed characteristics of the supersoft X-ray
sources in novae, and it is likely
to have another origin, possibly a hot spot on the WD surface. 

Did we observe IM Nor   too early, when
 the supersoft emitting hot WD atmosphere was
absorbed by the intrinsic 
column density of the ejecta? Even if the spectral
fits do not allow to rule this out completely, because of the
high value of N(H) for the first component of the
double Raymond-Smith model in Table 1, the optical spectrum
of the nova observed a month earlier shows that 
the ejecta were optically thin  5 months after the outburst. It
is very unlikely that we had significant intrinsic absorption even a  month
later. 
Therefore nuclear burning  switched off, or was 
in the process of switching off, six months after the 
outburst. It takes about a year
to terminate thermonuclear burning completely (Prialnik
2004, private communication). The mass burning rate for  stable
 hydrogen thermonuclear burning in the accreted shell is 
 a few 10$^{-7}$ M$_\odot$  year$^{-1}$, and the burnt mass
must be added to the non-burnt mass. There is in fact  
a minimum mass necessary to sustain shell hydrogen burning, 
which is M$_{\rm ex}\leq 10^{-6}$ M$_\odot$ for  
the expected WD mass in RN, M$_{\rm WD} \geq$ 1  M$_\odot$ (Fujimoto 1982).  
This implies that  the leftover mass after
about 80 years of recurrence time, does not exceed $\simeq
1.1-1.3 \times 10^{-6}$ M$_\odot$.
If 80 years is the average period between outburst, 
IM Nor  accretes 0.1 M$_\odot$ in about 10$^7$
years, 10 times longer than the typical time scale 
assumed by Della Valle \& Livio (1996) to accumulate this mass in a RN.
Since these authors found that even with an order of magnitude shorter
time scale  
RN do not contribute significantly to type Ia SN explosions,
at least assuming the Chandrasekhar mass model for the outburst,
RN like IM Nor, hitherto thought
to be the best candidate progenitor, appear to contribute at most in 
a negligible way to the type Ia SN rate. 
Even if some outbursts were missed notwithstanding the long decline, 
the conclusion does not change because with such
 little leftover mass, the interoutburst time would need to
 be of the order of one year to make IM Nor a ``statistically significant'' 
 SN Ia candidate.  We do not rule out that 
some RN may become type Ia SN,
 but our result strenghtens and makes more compelling the conclusion
of Della Valle \& Livio (1996):
the path of a RN to type Ia SN takes such a long time that RN 
cannot contribute significantly to the SN Ia rate. 
 At least another class of progenitor other than RN is needed to explain 
the observed type Ia SN rate. 

%
%

\acknowledgments

We are very grateful to H. Tananbaum for granting 
director's discretionary time for the first {\sl Chandra} observation. 
 We wish to thank B.W. Miller (Gemini Obs.) for his         
assistance in obtaining optical spectra.
CEW acknowledges support from NSF Grant AST02-05814 for ground-based
observations. S. Starrfield receives partial support
from NDF and NASA grants to ASU.



\clearpage

\begin{figure}
\begin{center}
\includegraphics[width=11.5cm,angle=-90]{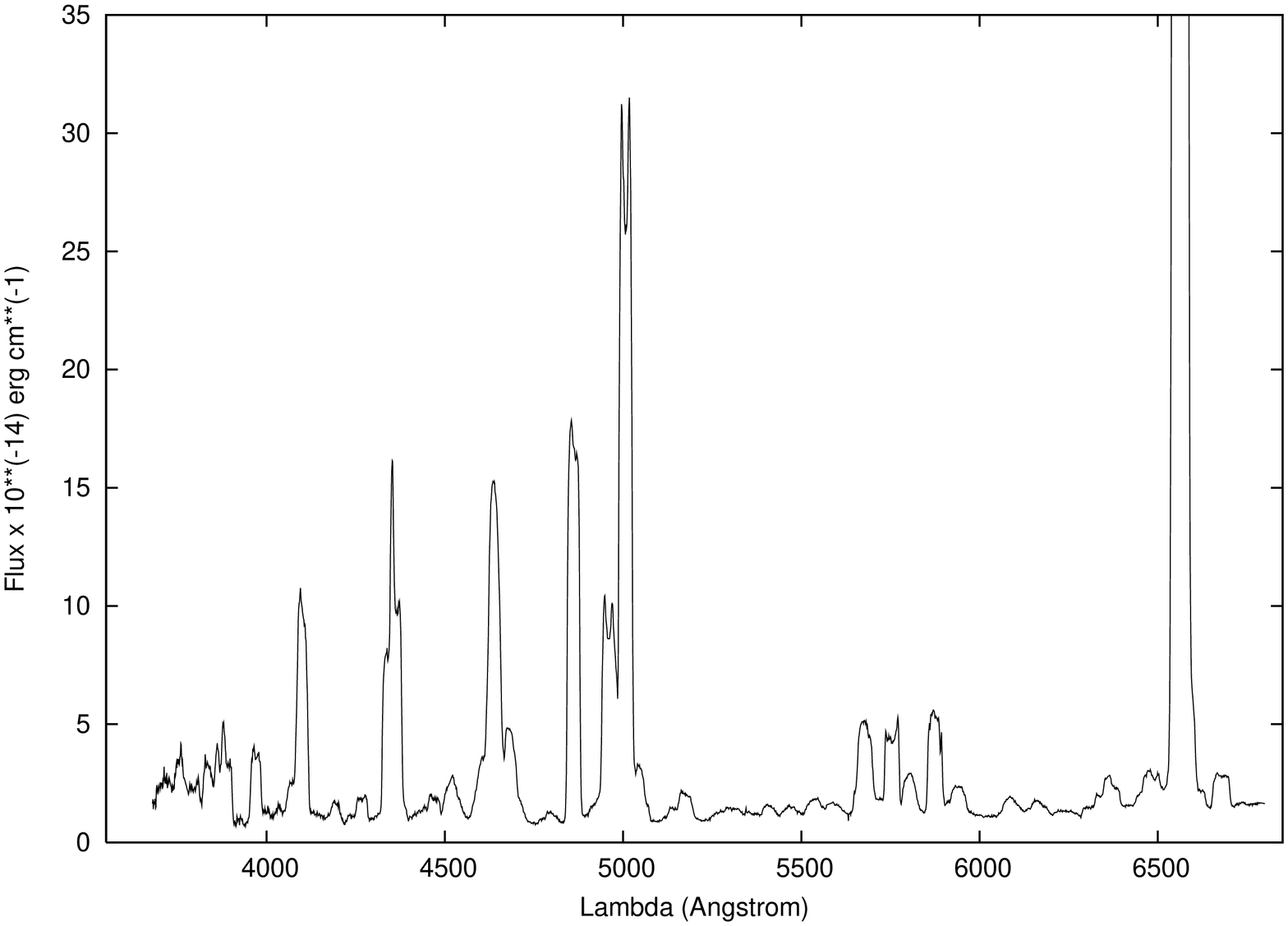}
\end{center}
\caption{ The optical spectrum of IM Nor observed with the CTIO Blanco 4m
telescope on 2002 April 29.4 UT}
\end{figure} 

\clearpage

\begin{figure}
\begin{center}
\includegraphics[width=10cm,angle=-90]{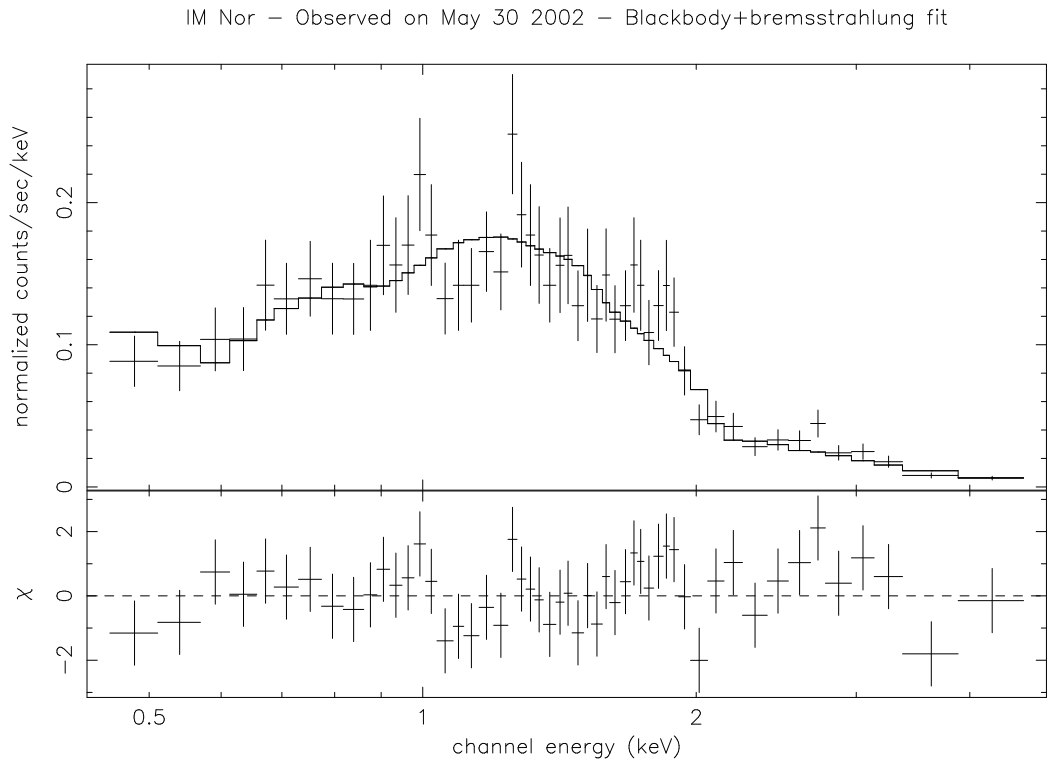}
\end{center}
\caption{ The spectrum of IM Nor observed with {\sl Chandra ACIS-S}
on 2002 May 30, fitted with 
the bremsstrahlung plus  blackbody model (see Table 1). }
\end{figure}

\clearpage

\begin{table}
\begin{center}
\caption{Parameters of the spectral fits: the plasma temperature kT and
black-body temperature T$_{\rm bb}$ are in units of keV, if not otherwise
specified. The column density N(H) is given in units of 10$^{21}$ cm$^{-2}$.
 The meaning of the N and k constants is specified in the text. In
{\sl VMEKAL(*)}, the abundances of several elements were allowed
to vary as also described in the text.   }
\begin{tabular}{|r|r|r|r|}
\tableline\tableline
Model & N(H) & Parameters &  $\chi^2$/dof \\
\tableline
power law & 3.6 & $\alpha$=2.34 & 1.50 \\
\tableline
bremsstrahlung & 2.6 & kT=2.4, N=7.74 $\times 10^{-4}$ & 1.20 \\
\tableline
Raymond-Smith & 1.6 & kT=4.7, N=1.24 $\times 10^{-3}$ & 2.15 \\
\tableline
MEKAL & 2.7 & Z$\simeq$0, kT=2.3, N=2.40 $\times 10^{-3}$ & 1.25 \\
\tableline
VMEKAL(*) & 3.7 & kT=1.6, N=4.30 $\times 10^{-4}$ & 1.06 \\
\tableline
VRAYMOND(*) & 3.0 & kT=1.8, N=4.16 $\times 10^{-4}$ & 1.03 \\
\tableline
blackbody+bremsstrahlung & 5 &  & 0.95 \\
blackbody  & & T$_{\rm bb}$=0.07, k=2.54 $\times 10^{-4}$ & \\
bremsstrahlung  & & kT=1.7, N=1.16 $\times 10^{-3}$ & \\
\tableline
bremsstrahlung+Gaussian & 2.5 & &  1.18 \\
bremsstrahlung & & kT=2.41, N=7.02 $\times 10^{-4}$ & \\
Gaussian & & E=1.94, $\sigma$=100 eV, N=6.7 $\times 10^{-6}$ & \\ 
\tableline
Raymond-Smith(1)+Raymond-Smith(2) & & & 1.10\\
 (1) & 32.0 & kT=339 eV, N=3.61 $\times 10^{-2}$ & \\
 (2) & 1.0 &  kT=5.5,  N=8.87 $\times 10^{-4}$  & \\ 
\tableline
\end{tabular}
\end{center}
\end{table}

\end{document}